\documentstyle[a4,12pt]{article}
\begin{document}
\hyphenation{ana-logue ana-lo-gous ana-lo-gous-ly ana-ly-sis}
\def\P{{\rm I\! P}}
\def\E{{\rm I\! E}}
\def\C{\hspace{2pt}{\sf l}\hspace{-5.2pt}{\rm C}}

\title{Comment on ``Hidden quantum nonlocality revealed by local filters''}

\author{Karin Berndl and Stefan Teufel\\ 
Mathematisches Institut der Universit\"{a}t M\"{u}nchen,\\
Theresienstra{\ss}e 39,
80333 M\"{u}nchen, Germany\\
e-mail: berndl@rz.mathematik.uni-muenchen.de;\\ md153ad@cip.mathematik.uni-muenchen.de}
\date{August 9, 1996}
\maketitle

In Section 3 of his paper \cite{Gisin}, Gisin argues that a ``careless
application of generalized quantum measurements can violate Bell's
inequality even for mixtures of product states.''  However, the observed
violation of the CHSH inequality is not in fact due to the application of
generalized quantum measurements, but rather to a misapplication of the
inequality itself --- to conditional expectations in which the conditioning
depends upon the measurements under consideration.

Consider the usual setup of quantum nonlocality arguments: a system 
consists of two widely separated subsystems, and in each of the subsystems 
one of two possible experiments $a,a'$ and $b,b'$, respectively, can be 
performed. The system is in a quantum state $\rho$ which is a density 
matrix on the Hilbert space ${\cal H}_1\otimes {\cal H}_2$. We shall denote 
by $O_a^i$ the positive operator on ${\cal H}_1$ giving via ${\rm tr}\, 
\rho \, (O_a^i\otimes I)$ the probability of obtaining the value $i$ for 
the measurement of $a$. Similarly $O_b^j$ denotes the positive operator on 
${\cal H}_2$ for the value $j$ of the measurement of $b$.\footnote{This 
covers measurements associated with positive operator valued (POV) measures, 
as well as the special case of measurements associated with self-adjoint 
operators, where $O_a^i$ will be the projection on the eigenspace of the 
eigenvalue $i$.}

A local hidden variables model for this setup consists of random variables 
$X_a$, $X_{a'}$, $X_b$, and $X_{b'}$ for the experiments under consideration 
on some probability space $(\Omega,\P)$ such that the joint distributions 
of the model reproduce the quantum joint distribution $P^\rho$
\begin{equation} \label{loc}
\P (X_a=i, X_b=j) = P^\rho (a=i,b=j) = {\rm tr}\, \rho \, (O_a^i\otimes O_b^j)
\end{equation}
for a joint measurement of $a$ and $b$, and similarly for the pairs $(a,b')$, 
$(a',b)$, and $(a',b')$. Thus we have for the expectation value of the product 
$a\cdot b$
\begin{equation}\label{expect}
E^\rho(a\cdot b) = \E( X_a X_b) \left(= \int_\Omega X_a (\omega) 
X_b (\omega) \, d\P(\omega) \right) ,
\end{equation} and similarly for the pairs $(a,b')$, $(a',b)$, and $(a',b')$.
Random variables $X_n$, $n=1\dots 4$ taking values in $[-1,1]$ satisfy the 
CHSH inequality \cite{CHSH,Bell}
\begin{equation} \label{CHSH}
\E(X_1 X_2) + \E(X_1 X_3) + \E (X_4 X_2) - \E (X_4X_{3}) \leq 2.
\end{equation} 
Thus (\ref{expect}) implies that if there is a local hidden variables model 
for quantum measurements taking values in $[-1,1]$ then 
\begin{equation} \label{qCHSH}
E^\rho(a\cdot b) + E^\rho(a\cdot b') + E^\rho(a'\cdot b) - E^\rho(a'\cdot b') \leq 2,
\end{equation} 
and the violation of (\ref{qCHSH}) proves that a local hidden variables model 
for the considered setup is impossible.
 
A quantum state $\rho$ that is a product state quite obviously allows for a
local hidden variables model reproducing the distributions of local
experiments, and thus this must also be true of a mixture of product
states. Moreover, this conclusion holds regardless of the nature of the
local experiments and in particular it does not matter whether these
experiments are described by standard observables represented by
self-adjoint operators or by generalized observables represented by
positive operator valued (POV) measures.\footnote{Gisin also states this in
the second sentence after Eqn.\ (12) in \cite{Gisin}. We have proven a more
general statement in \cite{unserpap}.}  Thus for quantum measurements with
results in $[-1,1]$, (\ref{qCHSH}) must be satisfied in a mixture of product
states. 

Nevertheless, Gisin presents POV's --- which may be regarded as
corresponding to the three possible outcomes $-1,0,1$ --- which apparently
yield a violation of (\ref{qCHSH}) even for a state which is a mixture of
product states. While the expectations that he considers only concern the
instances in which the particles both first pass through a filter, one
would expect the ensemble so defined to still be local and hence to still
satisfy (\ref{qCHSH}), even for generalized observables.

We thus must more carefully analyze the expectation values used by
Gisin. The expectation value of the product of $a$ and $b$ --- our $a$
corresponds to Gisin's $(\alpha, {\bf a})$, $b$ to $(\beta, {\bf b})$ etc.\
--- is given by
\begin{eqnarray} \label{cor}
E^\rho (a\cdot b) & = & \sum_{i,j\in\{ -1,0,1\} } ij \ P^\rho (a=i,b=j)\nonumber \\
& = & P^\rho( a = 1 , b = 1) + P^\rho( a = -1 , b = -1)\nonumber\\
& & - P^\rho( a = 1, b = -1) - P^\rho( a = -1, b = 1)
\end{eqnarray}
and analogously for $E^\rho (a\cdot b')$, $E^\rho (a'\cdot b)$, and 
$E^\rho (a'\cdot b')$. This equals the numerator in Gisin's Eqn.\ (10), and this 
quantity  cannot violate (\ref{qCHSH}) when calculated in a quantum state which is 
a mixture of product states. But the quantities for which Gisin shows that they 
can lead to a violation of (\ref{qCHSH}) --- Gisin's Eqn.\ (10) --- are not 
$E^\rho (a\cdot b)$ but the  conditional expectation values
\begin{eqnarray} \label{condcor}
E^\rho(a\cdot b| a \not=0 , b \not= 0) & = & \sum_{i,j\in\{ -1,0,1\} } ij \ 
P^\rho (a=i,b=j| a\neq 0, b\neq 0 )\nonumber \\
& = & \frac{E^\rho (a\cdot b)}{P^\rho (a \not= 0, b \not= 0)}
\end{eqnarray}
conditioned under ``both outcomes different from zero,'' {\it an event that
depends upon the the choice of experiments.\/} In a local hidden 
variables model these conditional expectations are represented by 
\begin{equation} \label{locrep}
E^\rho(a\cdot b| a \not=0 , b \not= 0) = \E( X_a X_b| X_a \not=0 , X_b \not= 0) 
= \frac{\E (X_a  X_b)}
{\P (X_a \neq 0, X_b \neq 0)}.
\end{equation}
Clearly, since for the different random variables $X_a$, $X_{a'}$, $X_b$,
and $X_{b'}$ these conditional expectations refer to different subensembles
of the original ensemble defined by $(\Omega,\P)$, in general the
conditional expectations need not satisfy (\ref{CHSH}), and thus the
violation of this inequality does not preclude the existence of a local
hidden variables model in this case.  In Gisin's example the subensembles
$(X_a \neq 0, X_b \neq 0)$, $(X_a \neq 0, X_{b'} \neq 0)$, etc., correspond
to the event that the photons pass the filters which are put in the
directions $({\bf a},{\bf b})$, $({\bf a},{\bf b'})$, etc., respectively,
and the problem arises simply --- as Gisin points out --- because the
``filters depend on the measured quantity'' which amounts to selecting
experiment-dependent subensembles.
 
Thus the results Gisin presents in Section 3 of \cite{Gisin}  are not at all  
``bizarre.'' Nor are they related to the application of POV measures. 
In fact, one can easily construct a similar situation with 3-valued standard 
observables (however, of course not in Hilbert space dimension 2): take 2 spin-1 
particles, ${\cal H}=\C^3\times \C^3$, consider the spin observables $J_\alpha$ in 
direction $\alpha$ in the $x$-$z$-plane
\[ J _\alpha = \left( \begin{array}{ccc} \cos\alpha & \frac {\sin\alpha}{\sqrt2} & 
0\\  \frac {\sin\alpha}{\sqrt2} & 0 &  \frac {\sin\alpha}{\sqrt2} \\
0 &  \frac {\sin\alpha}{\sqrt2} & -\cos\alpha \end{array} \right) ,\]
and take the state $\rho = \frac 12 P_{ (1, 0,0) \otimes (1,0,0)} + 
\frac 12 P_{ (\frac 12 ,\frac 1{\sqrt2} ,\frac 12)\otimes (\frac 12,
\frac 1{\sqrt 2},\frac 12)}$. Then the conditional 
expectations $\widetilde E^\rho (\alpha ,\beta) = E^\rho(J_\alpha 
\otimes J_{\beta}| J_\alpha \neq 0, J_{\beta}\neq 0)$ satisfy
$\widetilde E^\rho (\alpha ,\beta) + \widetilde E^\rho (\alpha ',\beta) +
\widetilde E^\rho (\alpha ,\beta ' ) - \widetilde E^\rho (\alpha ',\beta') = 
\frac{16}9 \sqrt 2 >2$ for the choice $\alpha = 0$, $\beta =\frac \pi 4$, 
$\alpha '= \frac \pi 2$,  $\beta ' = - \frac \pi 4$.
This, however, as explained above, tells nothing about nonlocality, i.e., 
the nonexistence of a local hidden variables model.

We thank Shelly Goldstein for valuable discussions.


\begin{thebibliography}{99}

\bibitem{Gisin} N. Gisin, Phys. Lett. A 210 (1996) 151.
\bibitem{CHSH} J. Clauser, M. Horne, A. Shimony, R. Holt, Phys. Rev. Lett. 
23 (1969) 880.
\bibitem{Bell}  J.S. Bell, {\it Speakable and Unspeakable in
Quantum Mechanics}\/  (Cambridge University Press, Cambridge, 1987), pp.\ 36.
\bibitem{unserpap} S. Teufel, K. Berndl, D. D\"urr, S. Goldstein, and N. Zangh\`\i, 
Locality and Causality in Hidden Variables Models of Quantum Theory, 
in preparation.

\end{thebibliography}
\end{document}